\documentclass[12pt]{iopart}
\usepackage{bm}
\usepackage{graphicx}
\usepackage{color}

\begin{document}

\title{Site dilution in SrRuO$_3$: Effects on structural and magnetic properties}

\author{Renu Gupta and A K Pramanik}
\address{School of Physical Sciences, Jawaharlal Nehru University, New Delhi - 110067, India}

\eads{\mailto{akpramanik@mail.jnu.ac.in}}

\begin{abstract}
We have investigated the effect of site dilution with substitution of nonmagnetic element in SrRu$_{1-x}$Ti$_x$O$_3$ ($x$ $\leq$ 0.7). The nature of ferromagnetic state in SrRuO$_3$ is believed to be of itinerant type with transition temperature $T_c$ $\sim$ 162 K. Crystallographically, SrRuO$_3$ has a distorted orthorhombic structure. Substitution of Ti$^{+4}$ (3$d^0$) for Ru$^{+4}$ (4$d^4$), however, does not introduce significant structural modification due to their matching ionic radii. This substitution, on the other hand, is expected to tune the electronic correlation effect and the $d$ electron density in system. With Ti substitution, we find that magnetic moment and Curie temperature decreases but $T_c$ remains unchanged which has been attributed to opposite tuning of electron correlation effect and density of states within framework of itinerant ferromagnetism. The estimated critical exponent ($\beta$) related to magnetization implies a mean-field type of magnetic nature in SrRuO$_3$. The value of $\beta$ further increases with $x$ which is understood from the dilution effect of magnetic lattice. The system evolves to exhibit Griffiths phase like behavior above $T_c$ which is usually realized in diluted ferromagnet following local moment model of magnetism. Our detail analysis of magnetization data indicates that magnetic state in SrRuO$_3$ has contribution from both itinerant and local moment model of magnetism.    
\end{abstract}



\maketitle
\section {Introduction}
Understanding the effect of disorder on ferromagnet (FM) is a long standing issue in condensed matter physics. Of particular interest is itinerant ferromagnet which is realized in the framework of Stoner criterion i.e., $UN(\epsilon_F$) $>$ 1, where $U$ is the intra-site Coulomb repulsion or electronic correlation energy and $N(\epsilon_F$) is the density of states (DOS) at Fermi level.\cite{moriya} In fact, based on $U$ the itinerant FMs are classified into weak and strong limit where the correlation effect is weak and strong, respectively. The prototype examples for weak itinerant FMs are ZrZn$_2$,\cite{wohlfarth} Ni$_3$Al,\cite{boer} etc. while Fe is a	well known itinerant FM with strong correlation effect.\cite{uhl} There are, however, many itinerant FMs where the strength of $U$ falls in intermediate range. Therefore, introducing disorder in form of chemical impurity, which is expected to tune either or both the $U$ and $N(\epsilon_F)$ parameters, appears to be an effective route to understand the magnetism of original system.

The 4$d$ based transition metal oxide SrRuO$_3$ is a commonly believed itinerant FM which has shown many interesting properties.\cite{koster} Usually, 4$d$ transition metals have an intermediate strength of $U$ and spin-orbit coupling (SOC) effect compared to its 3$d$ and 5$d$ counterparts which show prominent $U$ and SOC, respectively. The SrRuO$_3$ crystallizes in orthorhombic symmetry with distorted perovskite structure (GdFeO$_3$-type) where RuO$_6$ octahedra exhibits both tilt and rotation.\cite{cao} In spite of large volume of studies, the nature of magnetism as well as strength of $U$ in this material remains highly debated. This material has long-range ferromagnetic transition temperature $T_c$ $\sim$ 162 K. However, the magnetic moment even measured in high magnetic field (1.4 $\mu_B$/f.u.) turns out lower than its calculated spin-only value 2 $\mu_B$/f.u.\cite{cao} Electrical transport data show throughout metallic behavior though there is slope change in resistivity around $T_c$ due to reduced spin fluctuation.\cite{allen} The x-ray photoemission spectroscopy (XPS) study reports $U$ is significantly weak in SrRuO$_3$.\cite{kalo-xps} On the other hand, experimental studies, for instance, photoemission spectroscopy (PES),\cite{kim-xps,takizawa} angle-resolved photo emission spectroscopy (ARPES),\cite{shai} optical spectroscopy \cite{jeong} have shown non-negligible strength of $U$ which is in favor of local moment behavior. A recent band theory calculation employing combination of density functional theory (DFT) and dynamical mean-field theory (DMFT) has shown weak itinerant type FM behavior below $T_c$ and local residual magnetic moment behavior above $T_c$, indicating a dual presence of weak itinerant and local moment behavior in SrRuO$_3$.\cite{kim-dft} With this picture, it is required to understand the nature of magnetism in SrRuO$_3$ more clearly, and introducing the chemical impurity to tune $N(\epsilon_F$) and $U$ would be an efficient route in this regard. Recently, a drastic variation of $T_c$ with film thickness has been shown for SrRuO$_3$ which has been attributed to the change in $N(\epsilon_F$) in ultrathin films of this material.\cite{chang} 

In this study, we report an evolution of structural and magnetic properties in SrRu$_{1-x}$Ti$_x$O$_3$ ($x$ $\leq$ 0.7) where the substitution of nonmagnetic Ti$^{+4}$ (3$d^0$) for Ru$^{+4}$ (4$d^4$) amounts to dilution of magnetic network (Ru-O-Ru) in original system. There have been several experimental and theoretical studies\cite{kalo-xps,kim-xps,kim-mit,kalo-calc,bianchi,lin,abbate} investigating transport and electronic properties in SrRu$_{1-x}$Ti$_x$O$_3$ but the evolution magnetic properties has not been looked yet in details. The introduction of Ti$^{+4}$, on other hand, is less likely to induce any major structural modification as both the elements have very close ionic radii (Ru$^{+4}$ = 0.62 \AA and Ti$^{+4}$ = 0.605 \AA). Following this substitution of Ti$^{+4}$, one can expect an increase in $U$ and depletion of electrons in original system due to its 3$d^0$ character. These changes in $U$ and $N(\epsilon_F)$ will definitely influence the magnetic behavior according to itinerant model of FM. In fact, PES study has shown that along with a coherent peak there is a presence of incoherent peak below Fermi level in SrRuO$_3$, which implies a presence of non-negligible $U$.\cite{kim-xps} This study further shows that, with increasing $x$, the ratio between incoherent and coherent peaks increases and the DOS at Fermi level $N(\epsilon_F)$ depletes. Band structure calculation employing \textit{GGA + U} technique has also shown an increase of $U$ with Ti substitution.\cite{lin} On the other hand, following the picture of local spin model of FM the dilution of spin interaction along Ru-O-Ru bond would modify the $T_c$ as well as magnetic moment. This dilution may further induce Griffiths phase\cite{griffiths} behavior above $T_c$ as has been observed in Sr$_{1-x}$Ca$_x$RuO$_3$ where the dilution is realized due to suppression of Ru-O-Ru bond angle with Ca doping.\cite{jin}

\begin{figure}
	\centering
		\includegraphics[width=8cm]{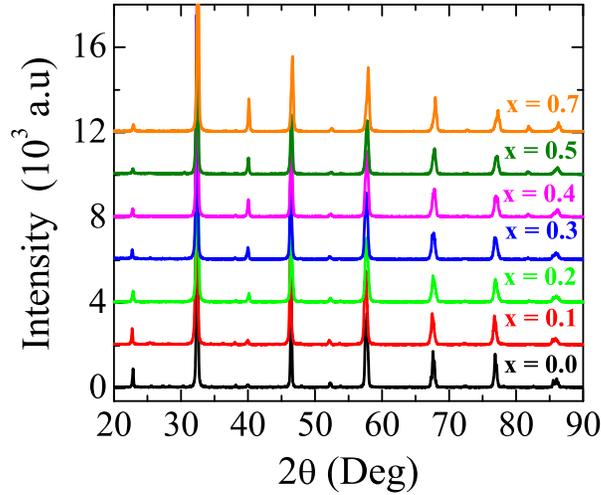}
	\caption{Room temperature XRD pattern have been shown for SrRu$_{1-x}$Ti$_x$O$_3$ series.}
	\label{fig:Fig1}
\end{figure}

Our studies show that original orthorhombic structure is retained with Ti substitution in SrRu$_{1-x}$Ti$_x$O$_3$ ($x$ up to 0.7) though structural parameters modify with $x$. While the $T_c$ remains almost unchanged in this series, we find both magnetic moment and Curie temperature decrease with $x$. For undoped SrRuO$_3$, the critical exponent ($\beta$) related to magnetization shows value close to the mean-field model which increases with Ti reaching about 1 for highest doped sample ($x$ = 0.7). Furthermore, the doped samples exhibit GP like behavior above $T_c$ where the behavior is prominently observed for $x$ $\geq$ 0.4. The analysis of thermal demagnetization data both itinerant and local-moment type of magnetism is present in SrRuO$_3$.           

\begin{figure}
	\centering
		\includegraphics[width=7cm]{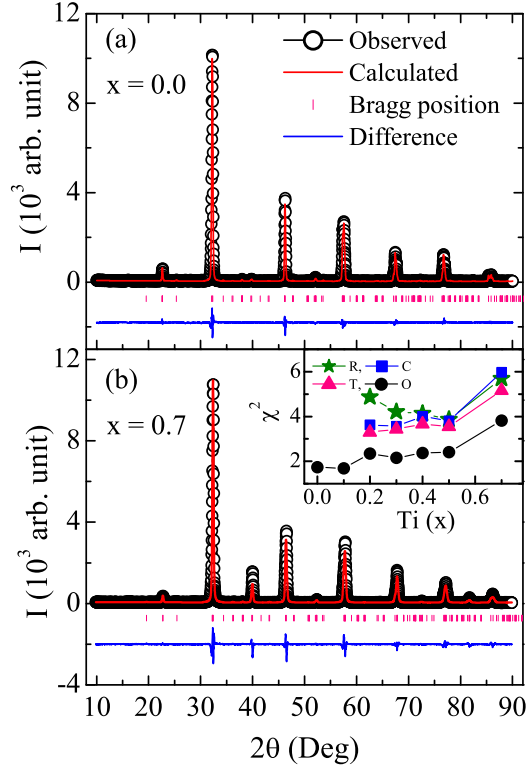}
	\caption{(a) The room temperature XRD pattern along with Rietveld refinement taking orthorhombic-\textit{Pbnm} structure have been shown for (a) SrRuO$_3$ and (b) SrRu$_{0.3}$Ti$_{0.7}$O$_3$. Inset in (b) shows $\chi^2$ value of Rietveld fitting for SrRu$_{1-x}$Ti$_x$O$_3$ series with rhombohedral (R), cubic (C), tetragonal (T) and orthorhombic (O) structure.}
	\label{fig:Fig2}
\end{figure}

\section {Experimental Details}
Polycrystalline samples of SrRu$_{1-x}$Ti$_x$O$_3$ with $x$ = 0.0, 0.1, 0.2, 0.3, 0.4, 0.5 and 0.7 have been prepared using solid state route. The ingredient powder materials of SrCO$_3$, RuO$_2$ and TiO$_2$ with phase purity greater than 99.99\% (Sigma-Aldrich) are taken in stoichiometric ratio and ground well. The mixed powders are heated in air at 1000 $^o$C for 24 h with for two times with an intermediate grinding. The calcined powders are then palletized and sintered at 1100$^o$C and 1150$^o$C for 36 h each time with an intermediate grinding. The phase purity of the samples is checked using powder x-ray diffraction (XRD) with a Rigaku MiniFlex diffractomer with CuK$_\alpha$ radiation. The data have been collected in 2$\theta$ range of 10-90$^o$ at a step of 0.02$^o$. To understand the structural evolution in this series, XRD data have been analyzed with Rietveld refinement program. DC magnetization ($M$) measurement is done using superconducting quantum interference device (SQUID) magnetometer (Quantum Design).

\begin{figure}
	\centering
		\includegraphics[width=8cm]{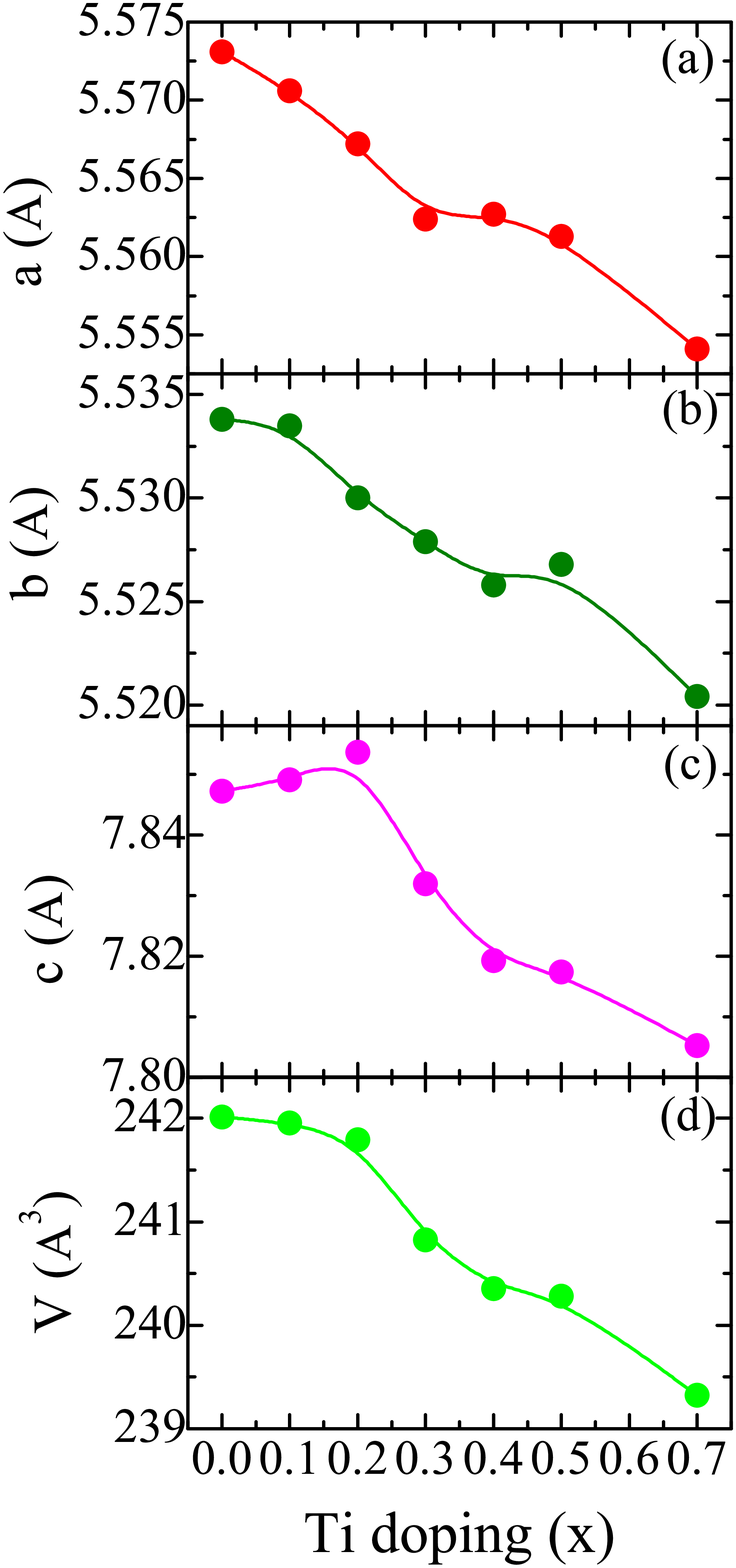}
	\caption{Unit cell parameters (a) $a$, (b) $b$, (c) $c$ and (d) volumn $V$ as determined form Rietveld analysis using room temperature XRD data have been shown with Ti concentration ($x$) for SrRu$_{1-x}$Ti$_x$O$_3$. Lines are guide to eyes.}
	\label{fig:Fig3}
\end{figure}

\section{Results and Discussions}
\subsection{Structural study}
Fig. 1 shows room temperature XRD pattern for SrRu$_{1-x}$Ti$_x$O$_3$ series with $x$ = 0.0, 0.1, 0.2, 0.3, 0.4, 0.5 and 0.7. The XRD pattern with diffraction peaks for the parent material ($x$ = 0.0) matches well with the reported study.\cite{jin} With this substitution of Ti for Ru, major structural modification is not expected considering their matching ionic radii (Ru$^{+4}$ = 0.62 \AA and Ti$^{+4}$ = 0.605 \AA). Indeed, Fig. 1 shows XRD pattern does not modify significantly with $x$ except a peak at 2$\theta$ $\sim$ 40$^{o}$, which is minimally present in $x$ = 0 compound, its intensity gradually increases with Ti. The XRD pattern has been analyzed using Rietveld refinement program.\cite{young} Fig. 2 shows representative XRD data along with Rietveld refinement for two end members of this series i.e., $x$ = 0.0 and 0.7. The refinement shows material with $x$ = 0 (SrRuO$_3$) crystallizes in orthorhombic structure with \textsl{Pbnm} symmetry (Fig. 2a). For the present SrRu$_{1-x}$Ti$_x$O$_3$ series, earlier studies have shown different evolution of structural phase with increasing amount of Ti. For instance, Cuffini \textit{et al.}\cite{cuffini} has shown that crystallographic structure changes from orthorhombic ($x$ = 0) to cubic phase with $x$ between 0.4 and 0.5. Bianchi \textit{et al.}\cite{bianchi} has shown that the system retains its orthorhombic structure till $x$ = 0.6 and above this it changes to tetragonal structure and finally changes to cubic structure for $x$ $>$ 0.7. Recently, Jang \textit{el al.}\cite{jang} has reported a single-phase orthorhombic structure up to $x$ = 0.1, then double-phase structure (orthorhombic and cubic) till $x$ = 0.5 and after that a single-phase cubic structure for $x$ $>$ 0.5. A single-phase orthorhombic structure has also been shown for $x$ up to 0.6.\cite{kalo-xps}

For all the materials in present SrRu$_{1-x}$Ti$_x$O$_3$ series with $x$ $\leq$ 0.7, we have tried to analyze the XRD data with all possible orthorhombic (\textit{Pbnm}), tetragonal (\textit{I4/mcm}), rhombohedral (\textit{R$\bar{3}$c}) and cubic (\textit{Pm$\bar{3}$m}) structure using Rietveld refine program. For SrRuO$_3$ (Fig. 2a), we find an orthorhombic structure with \textit{Pbnm} symmetry which is in agreement with majority of earlier studies.\cite{allen,cao,cuffini,bianchi,jang} With Ti substitution ($x$ up to 0.7), we find the same orthorhombic-\textit{Pbnm} structure is the best fitted one for whole series. While it has been previously shown that there is a phase transition from orthorhombic structure to more symmetric tetragonal or cubic structure above $x$ in range of 0.5 or 0.7,\cite{cuffini,bianchi} but we observe that original orthorhombic structure continues to be the best fitted structural phase till $x$ = 0.7 in this series. For completeness, statistical goodness of fit, which is gauged by $\chi^2$ value in Rietveld refinement, has been given for whole series as found with orthorhombic, tetragonal, rhombohedral and cubic structure in inset of Fig. 2b. As evident in figure, Rietveld refinement with orthorhombic structure gives lowest $\chi^2$ value for whole series with $x$ up to 0.7. Fig. 2b displays XRD pattern with Rietveld refinement with orthorhombic-\textit{Pbnm} structure for SrRu$_{0.3}$Ti$_{0.7}$O$_3$ showing reasonably good fitting of data. Here it can be mentioned that where other studies have reported structural phase transition from original orthorhombic structure around 50 to 70\% of Ti substitution in SrRu$_{1-x}$Ti$_x$O$_3$ series,\cite{cuffini,bianchi} our results demonstrating continuity of orthorhombic structure till 70\% of Ti concentration is in conformity with earlier results. Moreover, considering the matching ionic radii as well as ionic state of Ru and Ti, it is more likely that doped SrRu$_{1-x}$Ti$_x$O$_3$ materials will produce single and homogeneous structural phase rather than a situation of phase coexistence.    

Fig. 3 shows composition dependent evolution of lattice parameters i.e., $a$, $b$, $c$ and volume $V$ related to orthorhombic-\textit{Pbnm} phase for present SrRu$_{1-x}$Ti$_x$O$_3$ series. It is seen in figure that lattice parameters as well as volume decrease with Ti concentration, except the parameter $c$ which initially increases with $x$. The decrease of volume with Ti concentration can be explained with reduced ionic size of Ti$^{4+}$ compared to Ru$^{4+}$, and the similar behavior has also been observed by Cuffini \textit{et al.}\cite{cuffini} As expected, Ti substitution has not occurred no major structural modification as for the maximum doped ($x$ = 0.7) sample structural parameters (Fig. 3) modify only in range of (0.2 - 0.5)\%.  

\begin{figure}
	\centering
		\includegraphics[width=8cm]{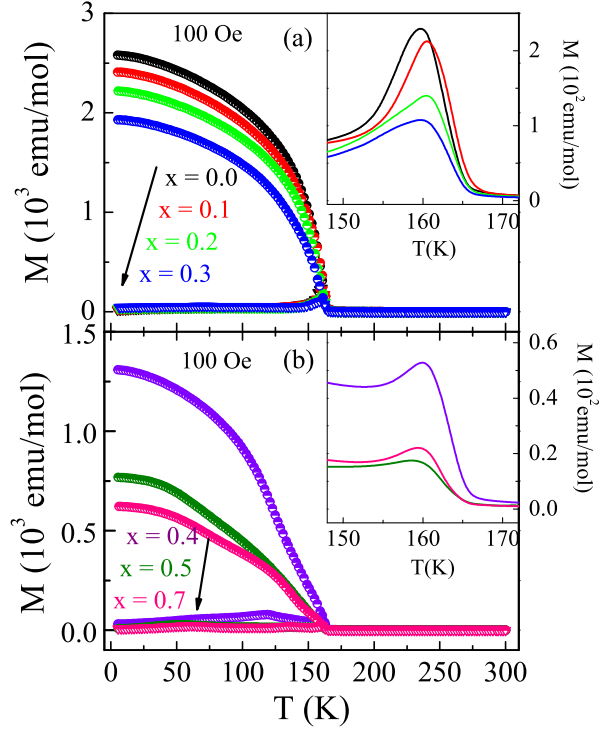}
	\caption{Temperature dependent magnetization data measured in 100 Oe following ZFC and FC protocol have been shown for SrRu$_{1-x}$Ti$_x$O$_3$ series with (a) $x$ = 0.0, 0.1, 0.2, 0.3 and (b) $x$ = 0.4, 0.5, 0.7. Insets show the zero field cooled magnetization data in expanded scale close to $T_c$.}
	\label{fig:Fig4}
\end{figure}

\subsection{Zero field cooled and field cooled magnetization data}
Temperature dependent magnetization data measured in 100 Oe magnetic field following zero field cooled (ZFC) and field cooled (FC) protocol for series SrRu$_{1-x}$Ti$_x$O$_3$ have been shown in Fig. 4. The Fig. 4a shows $M(T)$ for samples with $x$ = 0.0, 0.1, 0.2 and 0.3 and the Fig. 4b shows the same for $x$ = 0.4, 0.5, and 0.7. The inset of both the figures show ZFC magnetization data in limited temperature range. It is seen in Fig. 4a that ZFC and FC branches of magnetization data for parent compound with $x$ = 0.0 bifurcates at temperature $\sim$ 162 K. The inset of Fig. 4a shows $M_{ZFC}$ for $x$ = 0.0 shows a peak around this temperature. The large bifurcation between $M_{FC}$ and $M_{ZFC}$ indicates the material SrRuO$_3$ has large anisotropy.

With dilution of magnetic lattice in SrRuO$_3$ by substitution of nonmagnetic Ti$^{4+}$, the magnetization data in Fig. 4 is quite interesting. For samples with $x$ = 0.0, 0.1, 0.2 and 0.3, the bifurcation temperature between $M_{ZFC}$ and $M_{FC}$ as well as peak temperature in $M_{ZFC}$ remains almost same though the value of $M_{FC}$ decreases. The similar features are also observed for higher doped samples ($x$ = 0.4, 0.5 and 0.7) as evident in Fig. 4b. We have made an estimate of $T_c$ for this series from an inflection point in $M_{ZFC}(T)$ (using d$M$/dT plot) which is given in Table I as well as from fitting with Eq. 5 (discussed in section 3.4). It is rather intriguing that site dilution in terms of substitution of Ti$^{4+}$ at Ru$^{4+}$ site decreases the moment but the ferromagnetic ordering temperature $T_c$ appears to remain unchanged.     

\subsection{Thermal demagnetization study}
To understand the magnetic nature of SrRuO$_3$, we have analyzed the low temperature magnetization rather demagnetization data. Fig. 4 shows at low temperature $M_{FC}(T)$ continuously decreases with temperature. This thermal demagnetization can be explained both within scenario of localized as well as itinerant model of spin interaction. In localized model, the thermal demagnetization of $M(T)$ occurs due to excitation of spin-wave (SW) where the magnetization decreases following Bloch law as;\cite{kittel}

\begin{eqnarray}
	M(T) = M(0)\left[1 - BT^{3/2}\right]
\end{eqnarray}
  
where $M(0)$ is the magnetization at 0 K, and $B$ is the coefficients. The spin-wave stiffness constant $D$ can be calculated as;

\begin{eqnarray}
	D = \frac{k_B}{4\pi}\left[\frac{2.612g\mu_B}{M(0)\rho B}\right]^{2/3}
\end{eqnarray}

where $k_B$ is the Boltzmann constant and $\rho$ is the density of material. On the other hand, in itinerant or band model of magnetism the origin of magnetization is considered to arise due to displacement between spin-up and spin-down sub-band. The thermal demagnetization in this model is explained with the excitation of electron from one sub-band to another one. For strong itinerant type of FM, where one sub-band is completely filled and another one is partially filled, the single-particle excitation or thermal demagnetization is described as;

\begin{eqnarray}
	M(T) = M(0)\left[AT^{3/2}\exp(-\frac{\Delta}{k_BT})\right]
\end{eqnarray}
 
where $A$ is the coefficient, $k_B$ is the Boltzmann constant and $\Delta$ is the energy gap between the top of full sub-band and Fermi level. For the weak itinerant, where both the sub-bands are partially filled and have empty states at Fermi level, the single-particle excitation follows as;

\begin{eqnarray}
	M(T) = M(0)\left[AT^{2}\right]
\end{eqnarray}

where $A$ is the coefficient. However, it is possible to describe the nature of magnetism in a particular material using both localized and itinerant model simultaneously. 

\begin{figure}
	\centering
		\includegraphics[width=7cm]{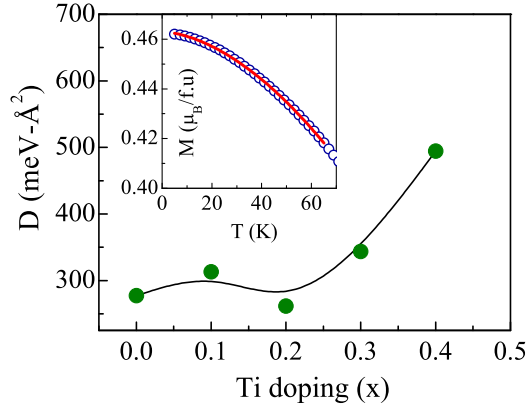}
	\caption{Spin-wave stiffness constant $D$ has been shown as a function of Ti doping concentration. Inset shows fitting of magnetization data using Eq. 1 and 4.}
	\label{fig:Fig8}
\end{figure}

In an aim to understand the nature of magnetism vis-$\grave{a}$-vis thermal demagnetization process in SrRu$_{1-x}$Ti$_x$O$_3$, we have analyzed the $M_{FC}(T)$ data (Fig. 4) with the above discussed models. The fitting with only Eq. 1 for spin-wave model has not given good fitting. Regarding single-particle model, the Eq. 3 can not be a valid model for thermal demagnetization as the spin polarization for SrRuO$_3$ is much lower than 100\%.\cite{nadgorny,pratap} Similarly, Eq. 4 alone did not give good result. Therefore, we have tried to fit the magnetization data taking combination of Eq. 1 and 4. The inset of Fig. 5 shows $M_{FC}(T)$ data along with fitting with combination of Eq. 1 and 4 i.e., with $M(T)$ = $M(0)\left[1 - BT^{3/2} - AT^2\right]$ up to temperature range 0.4 $T_c$ for SrRuO$_3$. We obtain fitting parameters as, M(0) = 0.463 $\mu_B$/f.u., B = 9.0 $\times$ 10$^{-5}$ K$^{-3/2}$ and A = 1.17 $\times$ 10$^{-5}$ K$^{-2}$. These obtained values of coefficient $B$ and $A$ match well with the earlier reported values.\cite{sow} Using Eq. 2, we have calculated spin wave stiffness constant $D$ = 264.9 meV$\AA^{2}$. Furthermore, exchange coupling constant ($J$) between nearest-neighbor magnetic atoms has been calculated to be 37.6 $k_BK$ following $B$ = (0.0587/$S$)$[(k_B)/(2SJ)]^{3/2}$ where $S (=1)$ is the localized atomic spin.\cite{kittel}

For the doped materials, we have analyzed demagnetization data and estimated stiffness constant $D$ following above mentioned procedure for $x$ up to 0.4. For the materials with $x$ $>$ 0.4, magnetic lattice is so diluted that analysis did not yield good result. The estimated $D$ has been shown in main panel of Fig. 5 which shows its value increases with $x$. Given that $T_c$ remains constant and also the number of nearest neighbor magnetic atoms decreases with replacing magnetic Ru$^{+4}$ by nonmagnetic Ti$^{+4}$, this increase of $D$ is quite interesting. Though it needs further investigation, we explain this increase of $D$ primarily due to the fact that Ti severs the Ru-O-Ru magnetic channel for propagation of spin-waves, hence the stiffness constant value increases.

\subsection{Nature of magnetic state and critical exponent for temperature dependent magnetization}
The nature of magnetic interaction in FM is usually characterized by set of critical exponents associated with different universality classes.\cite{stanley} This universality classes are decided by the dimensionality of lattice ($d$) and the dimensionality of spin system ($n$) where they do not depend on the microscopic details of the system. For present series of samples, we have extracted the the critical exponent $\beta$, which is related with the temperature dependent magnetization near $T_c$, and the critical temperature $T_c$ using following relation,

\begin{eqnarray}
	M = M_0(T_c - T)^\beta
\end{eqnarray}
 
The inset of Fig. 6 shows representative fitting of $M_{FC}$ data of SrRuO$_3$ near $T_c$ (up to $\sim$ 0.8$T_c$) using Eq. 5 where $M_0$, $T_c$ and $\beta$ have been kept as free fitting parameters. This fitting gives exponent $\beta$ = 0.48(2) and $T_c$ = 161.2(3) K. This value of $\beta$ is very close to the theoretically predicted value (0.5) for mean-field spin interaction model. Moreover, this value is consistent with the other report showing spin interaction in SrRuO$_3$ follows mean-field model.\cite{fuchs} The obtained $T_c$ is also close to the value estimated from temperature derivative of ZFC magnetization (Table I).  

\begin{figure}
	\centering
		\includegraphics[width=8cm]{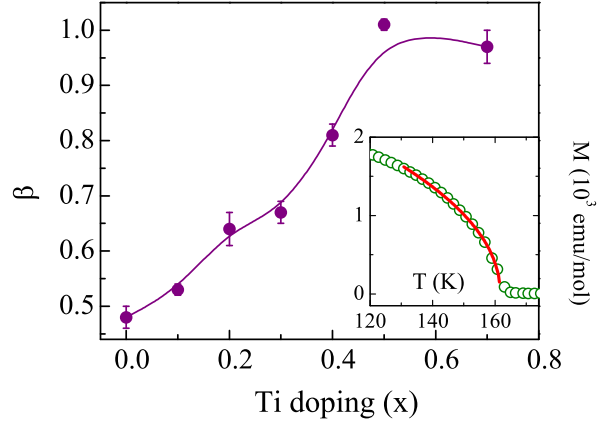}
	\caption{The evolution of critical exponent $\beta$ (Eq. 5) with $x$ has been shown for  SrRu$_{1-x}$Ti$_x$O$_3$ series. Inset shows field cooled magnetization data along with fitting due to Eq. 5.}
	\label{fig:Fig6}
\end{figure}

The main panel of Fig. 6 shows composition dependent evolution of $\beta$. The figure shows that value of $\beta$ increases with Ti substitution reaching close to 1.0 in highest doped material ($x$ = 0.7). These values of $\beta$ above 0.5, however, can not be ascribed to any of the known universality classes related to standard models. Nonetheless, this systematic evolution of exponent $\beta$ with Ti substitution is quite intriguing. The similar evolution of $\beta$ has also been observed for isovalent doped Sr$_{1-x}$Ca$_x$RuO$_3$.\cite{fuchs} In case of Sr$_{1-x}$Ca$_x$RuO$_3$, FM is weakened with Ca substitution and $T_c$ is completely suppressed at $x$ $\sim$ 0.7.\cite{cao,fuchs} For the doped materials, $T_c$ obtained from fitting with Eq. 5 even shows values remain close to that of parent material. The increase of exponent $\beta$ has been related to phase segregation between strongly coupled regimes surrounded by regimes with diluted FM spin-spin interaction.\cite{fuchs} This is supported by the fact that $\chi^{-1}(T)$ above $T_c$ in Sr$_{1-x}$Ca$_x$RuO$_3$ exhibit anomalous behavior showing Griffiths singularity which is characterized by preformed FM clusters in PM state.\cite{jin} It is worthy to note that with substitution of another isovalent ion Ba$^{2+}$, which has higher ionic radii than Sr$^{2+}$, in Sr$_{1-x}$Ba$_x$RuO$_3$ the exponent $\beta$ decreases and the end member BaRuO$_3$ shows exponents in agreement with 3D Heisenberg interaction where $\chi^{-1}(T)$ shows conventional Curie-Weiss type behavior ($T_c$ $\sim$ 60 K).\cite{jin,zhou-bro} In present SrRu$_{1-x}$Ti$_x$O$_3$, however, site dilution by Ti substitution results in a situation where the ferromagnetically aligned clusters surrounded by nonmagnetic matrix evolves and the size of those clusters increasingly decreases with substitution. This is evident with in Fig. 5 where the GP behavior is strengthened with $x$. It is interesting to see the similar type of evolution of critical exponent $\beta$ with the chemical substitution both at Sr- and Ru-site.  

\subsection{Griffiths phase behavior}    
Fig. 7 shows temperature dependence of inverse susceptibility [$\chi^{-1}$ = $(M/H)^{-1}$] for the selected samples with $x$ = 0.0, 0.3, 0.4, 0.5 and 0.7 for SrRu$_{1-x}$Ti$_x$O$_3$ series. The figure shows that $\chi^{-1}(T)$ shows fairly linear behavior at high temperature above $T_c$ following Curie-Weiss (CW) behavior (discussed later). However, the $\chi^{-1}(T)$ shows a sharp downfall immediately above $T_c$ for higher doped samples with $x$ $>$ 0.3 which indicates CW behavior is exactly not followed in this temperature regime. This sharp and sudden downfall in $\chi^{-1}(T)$ in the vicinity of $T_c$ is understood with the scenario of Griffiths phase (GP) behavior which is caused by phase inhomogeneity above $T_c$.\cite{griffiths} The GP was originally discussed for the diluted Ising ferromagnet with random distribution of nearest-neighbor exchange constant $J$ and 0 having probability $p$ and 1-$p$, respectively. It has been shown that for $p$ less than critical value $p_c$, long-range FM ordering can not survive in the system. For 1 $\geq$ $p$ $\geq$ $p_c$, system can have long-range magnetic ordering although transition temperature $T_c(p)$ is lower than that $T_c(p=1)$ of undiluted system. Conventionally, the $T_c(p=1)$ is called the Griffiths temperature $T_G$ which is determined as the temperature where $\chi^{-1}(T)$ starts to deviate from linear behavior in high temperature PM state. The temperature range between $T_G$ and $T_c(p)$ is called Griffiths phase regime where the system exhibit neither perfect FM ordering nor PM behavior. Rather, there exists ferromagnetically ordered finite size clusters embedded in PM background. Due to presence of clusters, magnetization shows nonanalytic behavior. As a result, susceptibility diverges which is demonstrated by sharp downturn in $\chi^{-1}(T)$.\cite{magen,psmo-gp}
  
\begin{figure}
	\centering
		\includegraphics[width=8.5cm]{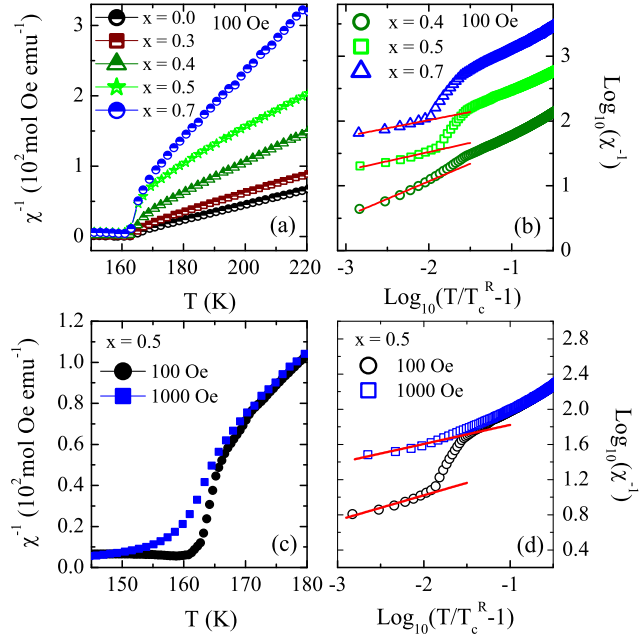}
	\caption{(a) shows temperature dependence of inverse magnetic susceptibility for SrRu$_{1-x}$Ti$_x$O$_3$ series with $x$ = 0.0, 0.3, 0.4, 0.5 and 0.7. (b) Inverse susceptibility has been plotted following Eq. 6 on $\log_{10}-\log_{10}$ scale with $x$ = 0.4, 0.5 and 0.7. (c) shows inverse susceptibility for $x$ = 0.5 sample in different magnetic fields. (d) shows same plotting as in (b) for $x$ = 0.5 sample with different magnetic fields. The straight lines in (b) and (d) are due to fitting with Eq. 6. }
	\label{fig:Fig7}
\end{figure}

The $\chi^{-1}(T)$ data in Fig. 7a shows a sudden and sharp downturn immediately above $T_c$ for higher doped samples which is prominent for $x$ $>$ 0.3. This behavior qualitatively shows GP like behavior which is reinforced with site dilution. In a quantitative manner, Griffiths singularity is characterized by following relation;\cite{neto}

\begin{eqnarray}
	\chi^{-1} = (T - T_c^R)^{1-\lambda}
\end{eqnarray}
                
where $\lambda$ is the exponent (0 $<$ $\lambda$ $\leq$ 1) and $T_c^R$ is the random critical temperature where the magnetic susceptibility tend to diverge. This Eq. 6 is a modified form of Curie-Weiss law for which $\lambda$ = 0. The finite value of $\lambda$ arises due to formation of FM clusters in PM regime, therefore, higher the value of $\lambda$ stronger is the GP behavior. Fig. 7b shows $\log_{10}-\log_{10}$ plot of Eq. 6 for samples with $x$ = 0.4, 0.5 and 0.7 where the GP behavior is prominently observed in Fig. 7a. The straight lines in low temperature GP regime (Fig. 7b) are due to fitting following Eq. 6 which is an indicative of Griffiths singularity in these materials. These straight line fittings give values of exponent $\lambda$ as 0.46, 0.72 and 0.75 and values of $T_c^R$ as 162.7, 160.8 and 161.3 K for $x$ = 0.4, 0.5 and 0.7, respectively. The obtained $T_c^R$ values are quite close to respective $T_c$ ($\sim$ 163 K) of these three materials. The $T_G$ which is usually determined as an onset temperature for downturn in $\chi^{-1}(T)$,\cite{psmo-gp} interestingly, has been found to be $\sim$ 166.8 K for all the sampleṣ with $x$ = 0.4, 0.5 and 0.7, in that sense GP regime ($T_G$ - $T_c$) is relatively narrow in present samples. The increasing value of exponent $\lambda$ with $x$ implies GP behavior is strengthened with site dilution effect.

We have further examined the GP behavior in varying magnetic fields. Fig. 7c shows $\chi^{-1}(T)$ for representative $x$ = 0.5 sample in magnetic field 100 and 1000 Oe. It is seen in figure that downturn in $\chi^{-1}$ across $T_c$ is reduced with increasing field. This can be explained as in higher fields the magnetic response from paramagnetic background becomes substantial to dominate over that from magnetic clusters. This leads to weakening of divergence of susceptibility and the $\chi^{-1}(T)$ tends to be linear. The similar behavior has also been observed in many other materials.\cite{magen,psmo-gp} Fig. 7d shows fitting of susceptibility data using Eq. 6 for $x$ = 0.5 sample with 100 and 1000 Oe magnetic field similar to Fig. 5b. From straight line fitting, we obtain exponent $\lambda$ = 0.72 and 0.78, and $T_c^R$ = 160.8 and 161.4 K for field 100 and 1000 Oe, respectively. The $T_c^R$ values are close to its $T_c$ ($\sim$ 163 K), and increase with magnetic field.

Here we mention that a close inspection in Fig. 7a reveals $\chi^{-1}$ for parent SrRuO$_3$ as well as for low doped samples ($x$ up to 0.3) show a weak downturn across $T_c$ which becomes more prominent above $x$ = 0.3. While this downturn in $\chi^{-1}$ for these low doped samples starts almost at same temperature of higher doped samples ($T_G$ $\sim$ 170 K, see Table I), the fitting with Eq. 6 for $\chi^{-1}$ data, however, does not give good result, probably due to weak nature of GP behavior. For this reason, we have not reported GP characteristic temperatures i.e., $T_c^R$ and $T_G$ for low doped samples in Table I. While $T_G$ is considered to be the FM ordering temperature of undiluted original system, the present $T_G$ ($\sim$ 170 K) being slightly higher than the $T_c$ ($\sim$ 163 K) of SrRuO$_3$ is not unusual as many parent materials are seen to show GP behavior where the essential disorder is realized coming from structural disorder or ionic mismatch.\cite{psmo-gp} For SrRuO$_3$, we observe very weak downturn in $\chi^{-1}$ because measurements are done in very low field (100 Oe) but the measurements done in high field (10000 Oe) do not show such behavior,\cite{jin} as the effect of magnetic field on GP behavior is seen in Fig. 7c. The appearance of GP behavior in present series is quite intriguing as the original SrRuO$_3$ is believed to be an itinerant type FM. However, signature of GP has been evidenced in isovalent doped Sr$_{1-x}$Ca$_x$RuO$_3$ which the authors have ascribed to the fact that suppression of Ru-O-Ru bond angle with Ca substitution dilutes FM interaction through Ru-O-Ru bonds, hence GP behavior is induced.\cite{jin} The GP picture is further supported by a recent band calculation using DFT and DMFT approach showing presence of both itinerant and local moment model of magnetism in SrRuO$_3$, particularly local type magnetic moment persists above $T_c$.\cite{kim-dft} In this scenario, we believe that GP behavior arises as a consequence of site dilution which renders small size ferromagnetically aligned clusters above $T_c$. The sizes of clusters are further reduced with Ti substitution which is evident from increasing value of exponent $\lambda$. Nonetheless, observed GP behavior brings out the local moment aspect of SrRuO$_3$.   

\begin{figure}
	\centering
		\includegraphics[width=7cm]{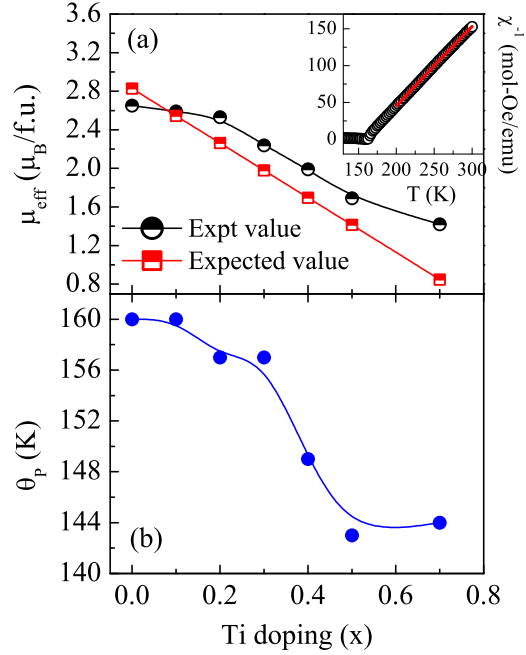}
	\caption{(a) The calculated and experimentally determined effective PM moment $\mu_{eff}$ have shown with $x$. Inset shows temperature dependent inverse susceptibility for SrRuO$_3$ and the straight line is fitting due to Eq. 7. (b) shows Curie temperature $\theta_P$ (Eq. 7) against Ti concentration.}
	\label{fig:Fig8}
\end{figure}

\subsection{High temperature paramagnetic state and Curie-Weiss behavior}
To understand the high temperature magnetic state, we have analyzed the magnetization data in terms of modified Curie-Weiss law;

\begin{eqnarray}
	\chi = \chi_0 + \frac{C}{T - \theta_P}
\end{eqnarray}

where $\chi$ is the magnetic susceptibility, $\chi_0$ is the temperature independent susceptibility, $C$ is the Curie constant and $\theta_P$ is the Curie temperature. The inset of Fig. 8a shows temperature dependence of inverse susceptibility $\chi^{-1}$ for SrRuO$_3$ in high temperature regime above $T_c$. The $\chi^{-1}(T)$ rather shows linear behavior in PM state. The Eq. 7 has been fitted with $\chi^{-1}(T)$ data in temperature range between 200 to 300 K which is shown as line in figure. The reasonably good fitting suggests magnetic susceptibility follows Curie-Weiss behavior. The fitting yields $\chi_0$ = 2.56 $\times$ 10$^{-4}$ emu mole$^{-1}$ Oe$^{-1}$, $C$ = 0.88 emu K mole$^{-1}$ Oe$^{-1}$ and $\theta_P$ = 160 K. Using this obtained Curie constant $C$, the effective PM moment $\mu_{eff}$ has been calculated to be 2.65 $\mu_B$/f.u. This experimentally obtained value for $\mu_{eff}$ is close to the expected value 2.83 $\mu_B$/f.u which has been calculated for spin-only moment g$\sqrt{S(S+1)} \mu_B$ with $S$ = 1. These values of $\theta_P$ and $\mu_{eff}$ reasonably agree with other studies.\cite{cao,kalo-xps} We find Curie-Weiss behavior is followed for whole series of samples at high temperatures (Fig. 7a). Fig. 8a shows composition dependent both experimentally observed and expected values of $\mu_{eff}$ for SrRu$_{1-x}$Ti$_x$O$_3$ series. As evident in figure, both values of $\mu_{eff}$ decreases with Ti substitution. While the expected value of $\mu_{eff}$ decreases linearly the observed $\mu_{eff}$ is not linear with $x$. Interestingly, the observed $\mu_{eff}$ is lower than the expected one for $x$ = 0, but it crosses over around $x$ = 0.1 and shows higher value with further increase of $x$. This can be explained from preformed FM clusters above $T_c$ giving a higher value of $\mu_{eff}$ which is also corroborated with the GP behavior as seen in Fig. 7. The value (160 K) of $\theta_P$ for SrRuO$_3$ turns out to be close to the bifurcation temperature (162 K) between $M_{ZFC}$ and $M_{FC}$ (Fig. 4). Moreover, the positive value of $\theta_P$ shows spin interaction is of ferromagnetic in nature. The $\theta_P$ as a function of Ti concentration is plotted in Fig. 8b which shows $\theta_P$ does not decrease significantly and remains positive with $x$ up to 0.7. This underlines the fact that though FM ordering is weakened but it survives with nonmagnetic Ti$^{+4}$ substitution as high as 70\% which is in sharp contrast with isovalent doped Sr$_{1-x}$Ca$_x$RuO$_3$ where FM ceases to exist beyond 70\% of Ca doping.\cite{cao}    

\begin{figure}
	\centering
		\includegraphics[width=8cm]{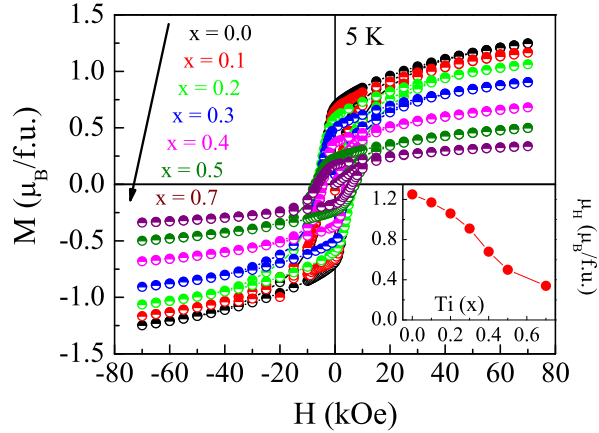}
		\caption{The isothermal magnetization data with field at 5 K are shown for SrRu$_{1-x}$Ti$_x$O$_3$ series. Inset shows value of moment at 70 kOe obtained from main panel with $x$.}
	\label{fig:Fig9}
\end{figure}

\subsection{Magnetic field dependent magnetization study}
The magnetic field dependent magnetization data collected at 5 K up to field 70 kOe have been shown in Fig 9. The undoped material SrRuO$_3$ shows a large hysteresis with coercive field $\sim$ 4800 Oe which matches with previous data.\cite{sow} At 70 kOe, the $M(H)$ data, however, do not saturate and continue to increase with smaller slope. At 70 kOe, we get moment $\mu_H$ about 1.25 $\mu_B$/f.u. which turns out smaller than the spin-only expected value 2 $\mu_B$/f.u. for total spin $S$ = 1. Here, it can be mentioned that moment in SrRuO$_3$ does not attain its expected full value in magnetic field as high as 30 Tesla.\cite{cao} This low value of moment is believed to arise due to itinerant magnetic character of SrRuO$_3$. With increasingly substitution of Ti, the moment decreases (inset of Fig. 9) though not linearly as expected from site dilution with nonmagnetic atom.     

\subsection{Analysis of Arrott plot and spontaneous magnetization}
We have further analyzed the $M(H)$ data in terms of Arrott plot which is plotting of $M^2$ vs $H/M$.\cite{arrott} The intercept of the slope in Arrott plot, which basically gives moment at $H$ = 0 and amounts to spontaneous moment ($M_s$), is very useful to understand the magnetic nature of a material. While negative intercept implies a non-ferromagnetic nature the positive value of intercept straightforwardly shows a ferromagnetic nature of material. Fig. 10 shows Arrott plot of $M(H)$ data taken at 5 K for whole series. As evident in figure, intercept of slopes taken in high field regime gives positive values for all the samples which suggests ferromagnetic ordering survives till highest concentration of Ti substitution (70 \%). Inset of Fig. 10 shows estimated $M_s$ against material composition ($x$) showing $M_s$ decreases with Ti following similar trend of $\mu_H$ in Fig. 9.

Now we check the Rhodes-Wohlfarth criterion\cite{wohl} which is usually used to distinguish between the localized and itinerant model of magnetization based on number of magnetic carriers per atom $q_c$ and $q_s$. The $q_c$ is estimated from effective PM moment $\mu_{eff}$ following $\mu_{eff}$ = $g\sqrt{S(S+1)}$, where $S$ is the effective spin per atom giving $S$ = $q_c/2$. The $q_s$ is obtained from low temperature saturation moment which is basically equals to $\mu_H$. In case of localized moment, $q_s$ or $\mu_H$ is realized from fully aligned moment, hence $q_c$/$q_s$ $\sim$ 1. For itinerant magnetism, $\mu_H$ is always lower than the fully aligned moment which results in $q_c$/$q_s$ $>$ 1. We calculate $q_c$ (from Fig. 8a) and $q_s$ (from Fig. 9) to be 1.82 and 1.25, respectively giving $q_c$/$q_s$ ratio 1.46. This ratio being higher than 1 obviously implies itinerant nature of magnetism in SrRuO$_3$, though based on all other studies it is inferred that magnetic state in SrRuO$_3$ is best explained by both itinerant and localized model of magnetism. 

\subsection{Behavior of transition temperature $T_c$ with site dilution}
So far we have seen that SrRuO$_3$ has both itinerant and local type of magnetism. With Ti substitution, magnetic moment and Curie temperature $\theta_P$ decreases but surprisingly $T_c$ is found remain unchanged. In following table we have given characteristic temperatures of present series.

\begin{table}
\centering
\caption{\label{tab:table 1} The characteristic temperatures $T_c$, $T_c^R$ and $T_G$ are given for SrRu$_{1-x}$Ti$_x$O$_3$ series.}

\begin{indented}
\item[]\begin{tabular}{ccccc}
\hline
Samples ($x$) &$T_c$ (K) &$T_c^R$ (K) &$T_G$ (K)\\
\hline
0.0 &162.7 & &\\
0.1 &162.6 & &\\
0.2 &162.9 & &\\
0.3 &162.6 & &\\ 
0.4 &162.6 &162.7 &170.6\\
0.5 &162.9 &160.7 &169.4\\
0.7 &162.9 &161.3 &169.9\\
\hline
\end{tabular}
\end{indented}
\end{table}

Following local moment model the magnetic moment, $\theta_P$ and $T_c$ are subjected to site dilution as it will weaken the FM state.\cite{kittel} In itinerant magnetism model, $T_c$ shows following functional dependence, 

\begin{eqnarray}
	T_c \propto \left[1 - \frac{1}{UN(\epsilon_F)}\right]^{1/2}
\end{eqnarray}

In present series of SrRu$_{1-x}$Ti$_x$O$_3$ materials, the replacement of Ru$^{+4}$ (4$d^4$) by Ti$^{+4}$ (3$d^0$) is expected to strengthen $U$ and weaken $N(\epsilon_F)$ owing to its 3$d^0$ character. Indeed, depletion of $N(\epsilon_F)$ and an increase of $U$ has been theoretically calculated and experimentally verified by using photoemission spectroscopy and x-ray absorption spectroscopy with an increase in Ti substitution.\cite{lin,kim-xps} We speculate that these opposite changes of $U$ and $N(\epsilon_F)$ will keep the $UN(\epsilon_F)$ term in Eq. 8 nearly constant which has resulted in $T_c$ without modification. While some of the experimental results such as, Curie-Weiss behavior, Griffiths phase behavior, spin-wave excitation can be explained with local model magnetism on the other hand, the low moment, single-particle excitation and unmodified $T_c$ can be understood form itinerant aspect magnetism in SrRuO$_3$. The findings in our present work are in conformity with recent theoretical calculations.\cite{kim-dft} We hope that our results will inspire further theoretical and experimental investigations using different kind of doping elements to understand this intriguing physics of itinerant ferromagnetism.

\begin{figure}
	\centering
		\includegraphics[width=8cm]{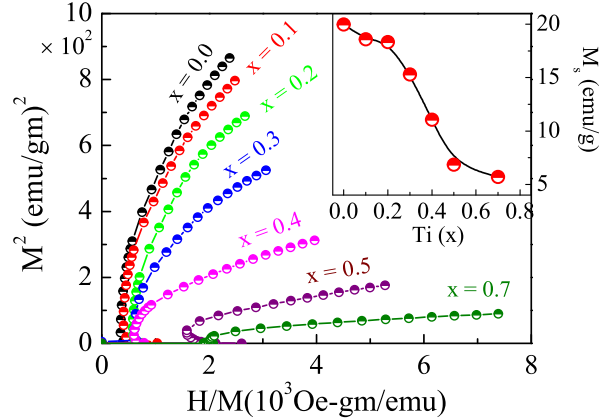}
	\caption{The $M(H)$ data plotted in form of $M^2$ vs $H/M$ (Arrott plot) for SrRu$_{1-x}$Ti$_x$O$_3$ series. Inset shows spontaneous moment $M_s$ as obtained from Arrott plot with Ti concentration.}
	\label{fig:Fig10}
\end{figure}

\section{Conclusion}
In summary, polycrystalline samples of SrRu$_{1-x}$Ti$_x$O$_3$ series with $x$ up to 0.7 have been prepared to understand the effect of site dilution on structural and magnetic properties. The parent material SrRuO$_3$ crystallizes in orthorhombic-\textit{Pbnm} structure having GdFeO$_3$ type crystallographic distortion where RuO$_6$ octahedra exhibits both tilt and rotation. With progressive Ti substitution, structural symmetry is retained though an evolution of structural parameters have been observed. Magnetic moment and Curie temperature ($\theta_P$) are observed to decrease with this dilution work. However, long-range magnetic ordering temperature $T_c$ remains unchanged which is understood from the opposite tuning of electron correlation effect and DOS with Ti substitution within model of itinerant magnetism. The estimated critical exponent $\beta$ (0.48) shows nature of magnetism in SrRuO$_3$ is of mean-field type, and increase of $\beta$ with $x$ has been ascribed to development of FM clusters with Ti substitution. Interestingly, similar to isovalent doped Sr$_{1-x}$Ca$_x$RuO$_3$, these materials exhibit Griffiths phase like behavior in higher doped samples which is again believed to arise from clustering effect above $T_c$ and represents the local moment picture of magnetism in SrRuO$_3$. Analysis of low temperature thermal demagnetization data is in favor of dual presence of itinerant and local moment in SrRuO$_3$ in conformity with recent theoretical calculation.

\section{Acknowledgment}   
We acknowledge UGC-DAE CSR, Indore and Alok Banerjee for the magnetization measurements. We are thankful to Kranti Kumar for the help in magnetization measurements. RG acknowledges UGC, India for BSR fellowship.

\end{document}